\documentclass[12pt,preprint]{aastex}

\shorttitle{Infrared Excess in the Be Star Delta Scorpii}
\shortauthors{Halonen et al.}

\begin{document}

\title{Infrared Excess in the Be Star $\delta$ Scorpii}

\author{R. J. Halonen, C. E. Jones and T. A. A. Sigut}
\affil{Physics and Astronomy Department, The University of Western Ontario,
    London, ON, N6A 3K7}
\email{rhalonen@uwo.ca   cjones@astro.uwo.ca   asigut@astro.uwo.ca}

\and

\author{R. T. Zavala}
\affil{United States Naval Observatory, Flagstaff Station, Flagstaff, AZ;}
\email{bzavala@nofs.navy.mil}

\and

\author{C. Tycner}
\affil{Department of Physics, Central Michigan University, Mt. Pleasant, MI}
\email{c.tycner@cmich.edu}

\and

\author{S. E. Levine, C. B. Luginbuhl and F. J. Vrba}
\affil{United States Naval Observatory, Flagstaff Station, Flagstaff, AZ;}

\and

\author{N. Vlieg}
\affil{Materials Science and Engineering Department, University of Arizona, Tucson, AZ}

\begin{abstract}
We present infrared photometric observations of the Be binary system $\delta$ Scorpii obtained in 2006. The J,H and K magnitudes are the same within the errors compared to observations taken 10 months earlier. We derive the infrared excess from the observation and compare this to the color excess predicted by a radiative equilibrium model of the primary star and its circumstellar disk. We use a non-LTE computational code to model the gaseous envelope concentrated in the star's equatorial plane and calculate the expected spectral energy distribution and H$\alpha$ emission profile of the star with its circumstellar disk. Using the observed infrared excess of $\delta$ Sco, as well as H$\alpha$ spectroscopy bracketing the IR observations in time, we place constraints on the radial density distribution in the circumstellar disk. Because the disk exhibits variability in its density distribution, this work will be helpful in understanding its dynamics. 

\end{abstract}


\section{Introduction}

Be stars are massive, rapidly rotating stars that exhibit distinct characteristics due to the presence of a thin disk of gas concentrated in the stars' equatorial plane. The classical definition of a Be star is that of a main or near-main sequence B star whose spectrum has, or had at some point in time, conspicuous Balmer emission lines \citep{col87}. In addition to this prominent emission line spectrum, important observational features of these rapid rotators include a partial polarization of radiated light and a continuum excess in the radio and infrared (IR) spectrum. These distinct observational characteristics of many Be stars are variable on timescales of weeks to decades, due primarily to changes in the physical conditions in the circumstellar disk. The preceding description limits classical Be stars to intermediate mass stars surrounded by gaseous envelopes and excludes related stars such as the Herbig Ae/Be stars which are pre-main sequence objects surrounded by dusty disks. 

The contribution of radiation from the circumstellar gas to the Be stellar continuum is significant. Observations of Be stars at IR wavelengths show a much brighter continuum spectra than would be expected from typical stellar photospheric emission. The explanation for this IR excess is that the continuum of the central star in the IR is dominated by free-free and free-bound emission from the equatorial disk \citep{geh74}. 

While evidence suggests that the gaseous, equatorial disk is formed from material released from the star, the physical processes that govern the formation of the disk are not well understood \citep{por03}, and are the subject of persistent study. The first viable models predicted an equatorial mass-loss disk triggered by the near-critical rotation of the central star \citep{str31}. While there is general agreement that the rapid rotation of the central star is fundamental to the disk formation mechanism, its proximity to the critical value remains a contentious topic \citep{tow04}. Other models of stellar mass ejection suggest that the gas outflow could be due to radiatively driven winds, possibly triggered by magnetic fields \citep{por97}, or due to pulsations in the stellar photosphere \citep{riv01}. Recently, the viscous decretion disk model pioneered by \citet{lee91} has successfully reproduced several key observables, including the IR excess, from Be star disks \citep{por99}. 

The Be star $\delta$ Sco is of particular interest to those studying the mechanisms of disk formation. In the summer of 2000, the star began exhibiting line emission in its radiated spectrum \citep{ote01}. Spectroscopic monitoring of the star over the following two years confirmed a continuous strengthening of the emission lines due to the steady development of a circumstellar disk \citep{mir03}. In addition, the brightness of the star has varied considerably, increasing by as much as 0.7 mag in the V band \citep{gan02}. Because $\delta$ Sco presents a unique opportunity of observing a Be star in an active disk formation phase, it is of particular interest for detailed astrophysical studies.

In this paper, we present unique infrared observations of the star $\delta$ Sco and we compare the observed IR excess to that predicted by our theoretical models. Our theoretical models are generated by a non-LTE code that enforces radiative equilibrium by a self-consistent disk thermal structure
and includes a solar-type chemical composition. For this investigation, following previous work to match the IR excess \citep{wat86, cot87, wat87}, we have adopted a simple density distribution which falls off as a power-law with increasing distance from the central star. Our models are also constrained by comparison of predicted H$\alpha$ profiles with contemporaneous observations. It should be noted that \citet{car06} reported that $\delta$ Sco's density distribution during this active stage of disk growth seems to deviate from a simple power law. However, as pointed out by \citet{oka07}, in order to assess the validity of the viscous decretion model it is important to carefully monitor the IR excess of Be stars to determine if the density index changes with time. Ascertaining a reasonable average value of the index at the time of IR observations, in conjunction with other work, will be an important test of the viscous disk model. In the following section, we discuss our photometric observations of $\delta$ Sco. In section 3, we present our models for the circumstellar disk around the studied star. Lastly, we summarize the results of our work in section 4 and discuss the implications of our findings.

\section{Observations}

\begin{figure}
\plotone{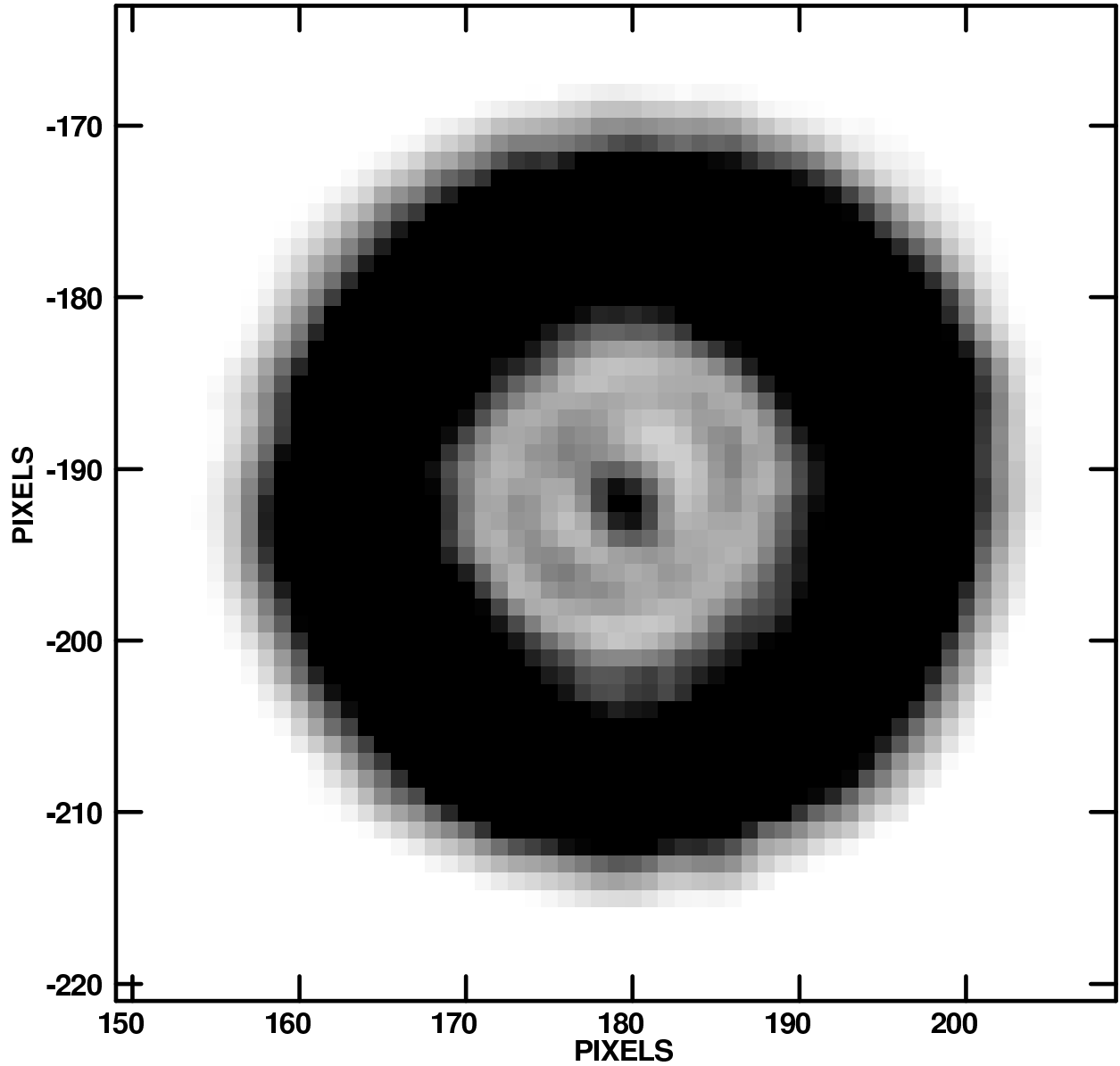}
\caption{Greyscale defoucsed K band image of $\delta$ Sco obtained on 2006 May 12. 
Although the maximum DN is greater than 130,000 the greyscale range is 
limited to 5,000 to 25,000DN to emphasize the donut-shaped image. The full width 
of the outer ring measured on several images is 38 pixels or 14''. The 
central peak is known as Poisson's spot and arises from the diffraction pattern 
from the secondary. Pixel values are relative to the array center. \label{donut}}
\end{figure}

$\delta$ Sco is a relatively bright \citep[K=2.61 $\pm$ 0.07 mag.,][]{tmss} star and presents 
difficulties for observations with anything but small aperture telescopes with modern array detectors. 
The telescope and detector used were the 61'' (1.55 m) Kaj Strand astrometric reflector 
and the Astrocam infrared array camera. The Strand reflector is an F/9.8 folded Newtonian 
with a plate scale of 13.5''mm$^{-1}$ \citep{strand}. Astrocam \citep{acam} contains a 1024 $\times$ 
1024 InSb array with a minimum readout time with double correlated sampling for the 
full array of 0.19 seconds. Astrocam pixels are 27 $\mu$ square providing 0.37 arcsec 
sampling with 1:1 reimaging of the focal plane. Observations were made on 2006 May 10, 11 
and 12. The first night was used as an 
engineering test run to determine the best strategy for observing $\delta$ Sco. 
Of the following nights only 2006 May 12 was deemed photometric. Initial tests showed that 
even with the minimum readout time Astrocam images of $\delta$ Sco were saturated. A +10 mm 
defocus was applied which produced the necessary donut shaped images (Fig.~\ref{donut}) to 
avoid saturation with 0.20 sec exposures and 50 co-adds in all filters (J,H,K).
The full width of the donuts was measured in several images and found to be 38 pixels, or 1.026 mm.
This is in excellent agreement with the geometrical optics expectation of 1.0204 mm 
for the Strand telescope. The central spot visible in Fig.~\ref{donut} is known as 
Poisson's spot and can be understood from diffraction (see Hecht and Zajac 1979 
for the source of this ironic name). All data and calibration images were obtained with the same +10mm 
defocus. We elected to observe both $\delta$ Sco and the photometric standards with this same 
defocus to simplify our analysis. Standard stars were chosen from \citet{g03} and are 
listed in Table~\ref{stndrd} with their K magnitudes, $J - H$ and $H - K$ colors, exposure times 
and number of co-adds which were the same for all filters. 

\begin{figure}
\plottwo{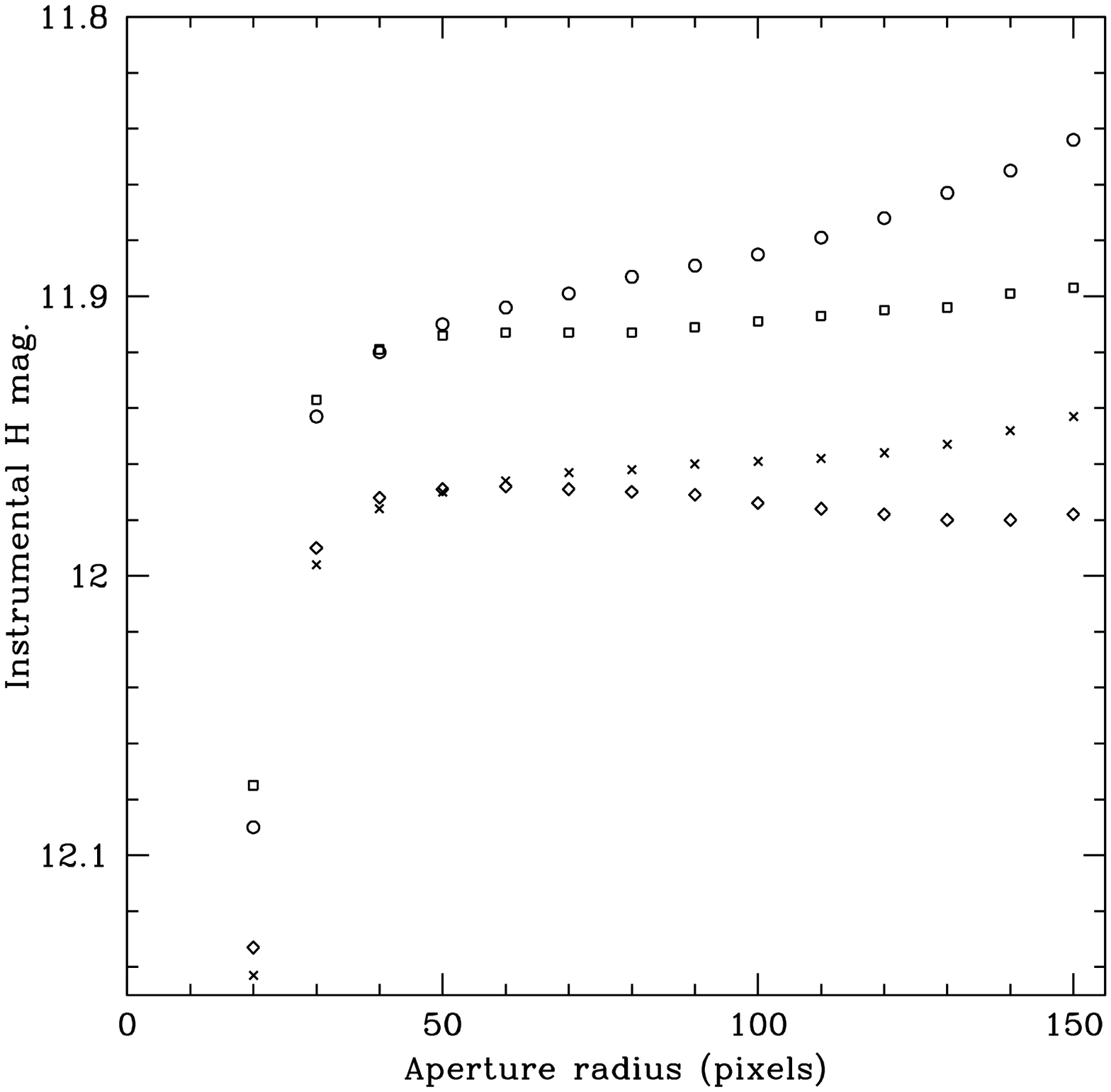}{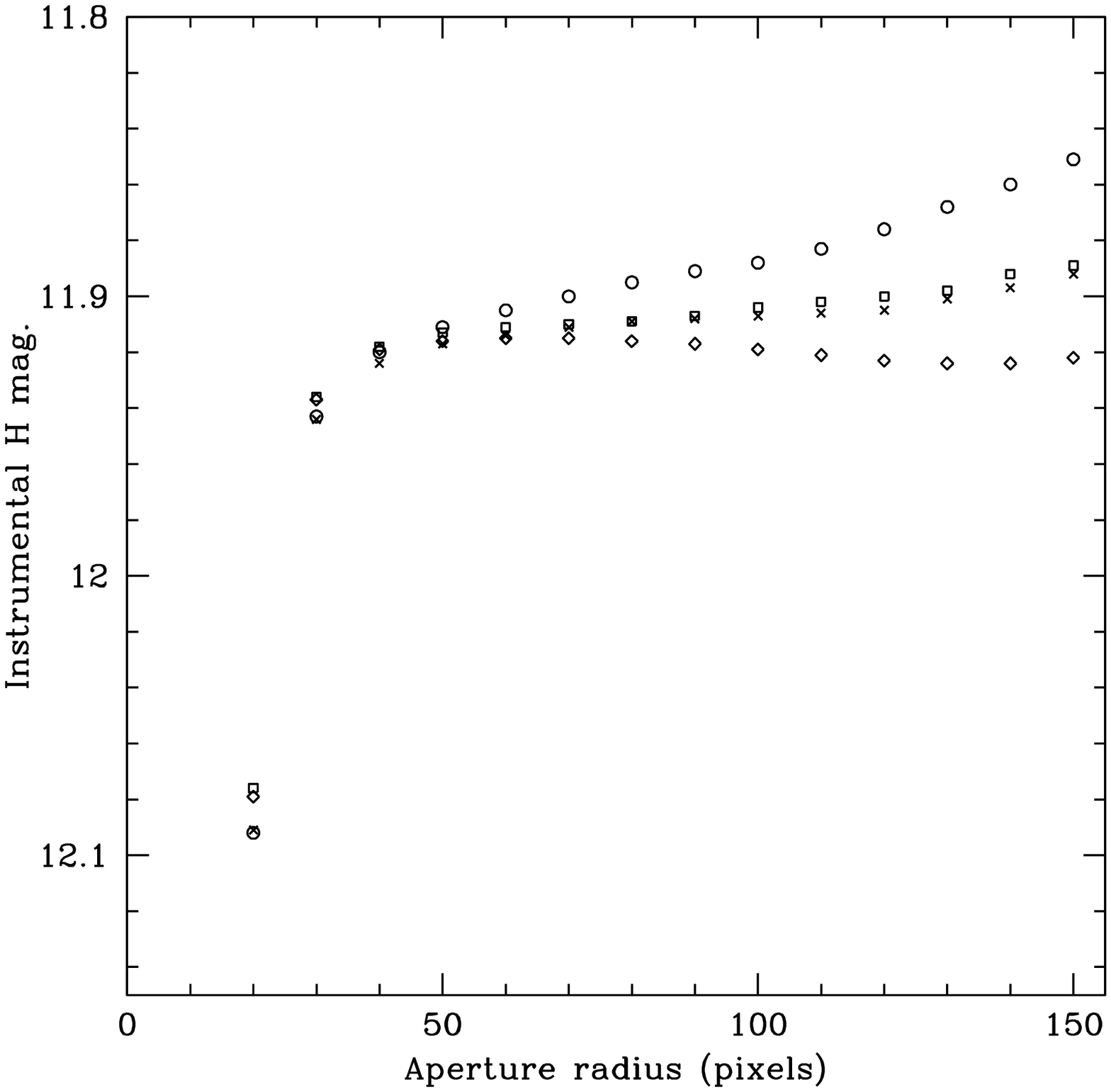}
\caption{Curves of growth in the H filter for the standard star HD 161961. Left 
panel is before the 5\% correction described in \S2 was applied to the lower 
half of the array. Right panel shows the same but after the 5\% correction. 
Error bars are comparable to the size of the points and are determined from 
the photon statistics. Different symbols correspond to one of the 4 dithered images. 
\label{cog}}
\end{figure}

Data reduction and image processing were performed with IRAF \citep{iraf}. Individual frames 
were linearized following a procedure described by \citet{lug98} and then flat-fielded. All stars
were observed with a 4 position dither pattern and these images were used to construct sky images 
for subtraction. A bad pixel mask was created and used to correct the images. After this process was 
performed each star was identified in the image and coordinate lists prepared. Aperture photometry 
was performed to produce curves of growth to determine the best parameters for aperture photometry. 
At this point we noticed that the 4 dithered images generally produced offset pairs of curves of 
growth. This offset was at the $\approx$ 5\% level and we attribute this to the 
different amplifiers used on Astrocam. This offset was always visible between the top and 
bottom halves of the array. We increased the counts in pixels below the 
camera midline by 5\% to compensate for this offset; Fig.~\ref{cog} shows 
the curves of growth before and after the 5\% correction was applied. For an aperture 
radius greater than 50 pixels we generally saw a divergence 
in the curves of growth and we therefore selected 50 pixels as the photometric 
aperture. We attribute this divergence to the difficulty in performing an accurate 
sky subtraction across large areas of the array and believe this ultimately limits 
our photometry.

A final factor to contend with were some cosmetically poor areas of the array. 
These areas had very different noise characteristics from the rest of the array. 
There was a significant positive bias in the noise which artificially elevated 
the count levels. The exact amount of the increase varied as a function of position in these  
roughly rectangular shaped areas. As this was very difficult to calibrate out we elected 
to disregard images for which any portion of the photometric aperture and sky annulus 
fell within these areas. Stars could be artificially brightened by up to 
0.2 magnitudes if they fell within these areas. This restriction eliminated 14 out 
of 96 standard star frames and 2 out of 48 $\delta$ Sco frames.

\begin{figure}
\epsscale{1.0}
\plotone{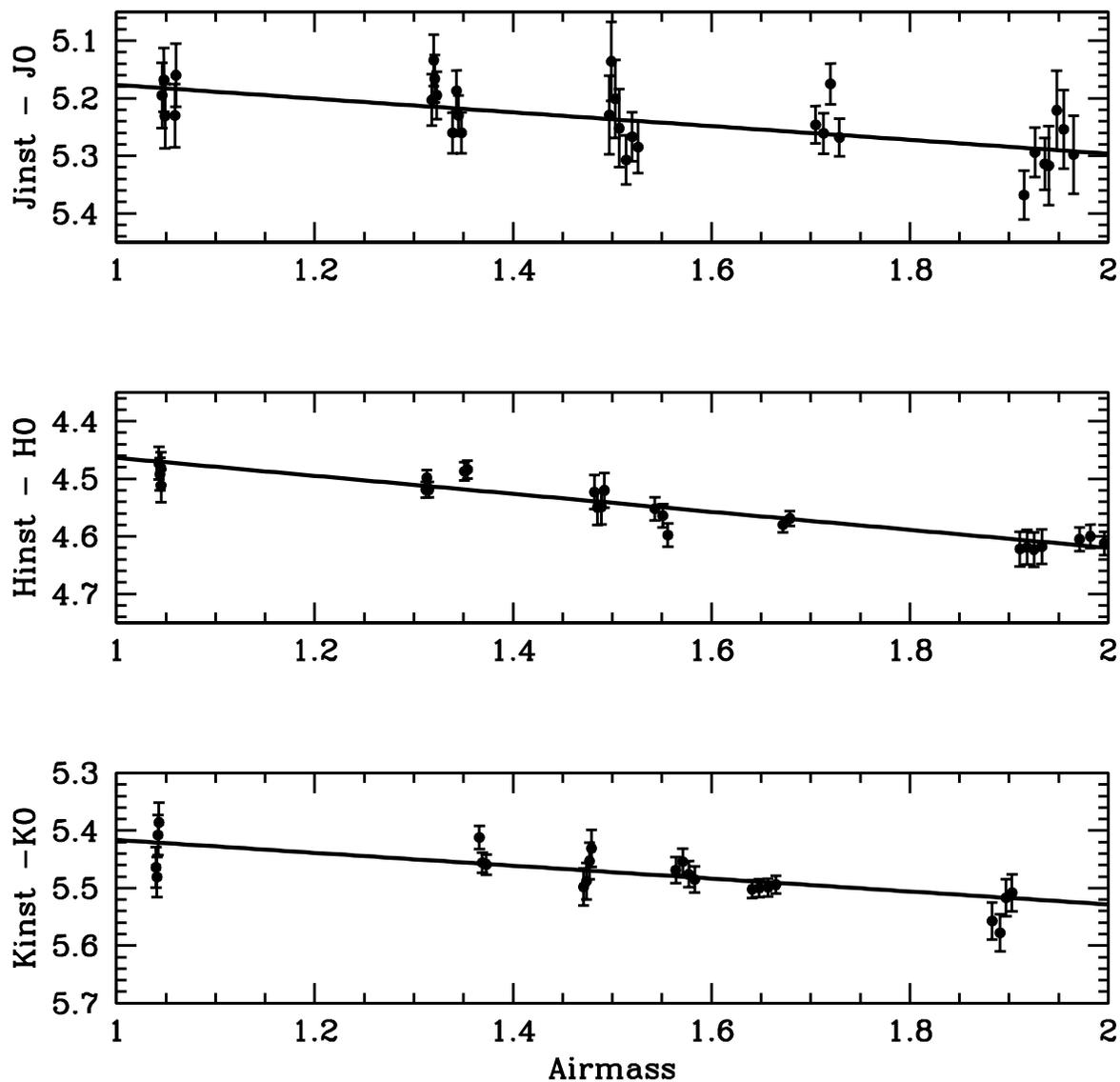}
\caption{Plots of fits for extinction (top=J, middle=H, bottom=K) for the 
standard stars observed on 2006 May 12. The magnitude differences on the 
y-axes are the instrumental magnitude  minus the  magnitude 
from \citet{g03} in the indicated filter. These fits exclude points which lay 
within the cosmetically poor regions discussed in \S2. Error bars have been 
increased from the photometric error value as discussed in \S2 and listed in 
Table~\ref{phot}. \label{ext}}
\end{figure}

\begin{figure}
\epsscale{1.0}
\plotone{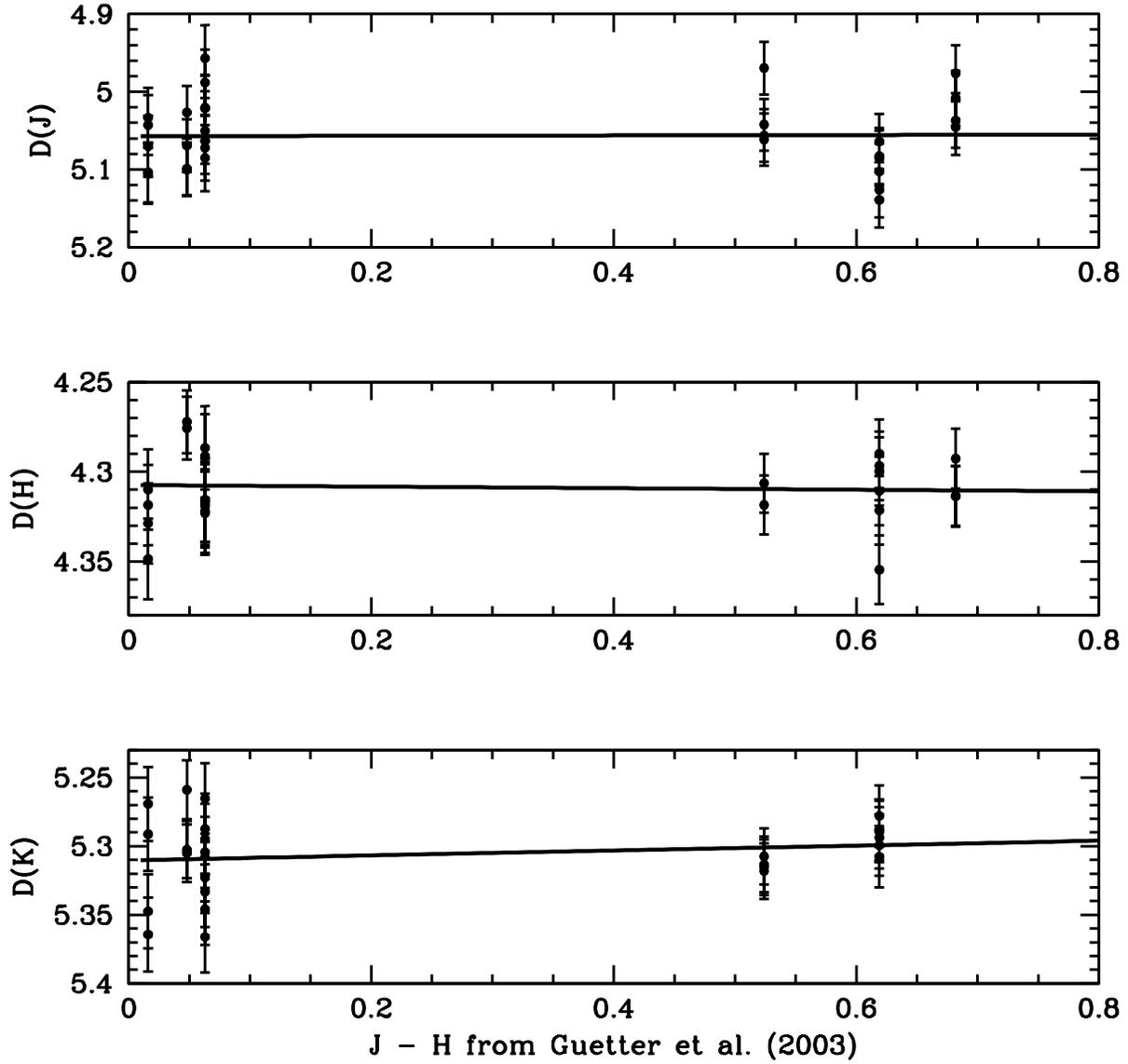}
\caption{Plots of magnitude differences of Fig.~\ref{ext} corrected for 
extinction using Columns 3 and 4 of Table~\ref{phot} as a function of color 
of the standard stars. Lines are plotted using color term results from Table~\ref{phot}. 
\label{color}}
\end{figure}

The aperture photometry results were exported to custom software written 
in IDL. Fits for the extinction were made in each filter and are shown 
in Fig.~\ref{ext}. The initial formal errors were determined by adding in quadrature 
the error in the instrumental magnitudes determined from photon statistics and those 
from \citet{g03}. These initial fits yielded large reduced $\chi^2$. We elected 
to increase the error bars by a factor of the square root of 
the reduced $\chi^2$ and these errors are shown in Fig.~\ref{ext}. 
We believe this increase in the errors is needed to properly account 
for the errors in estimating the sky subtraction across large areas 
of the array at the four dither positions. We corrected the standard star 
results for extinction and then tested for the presence of any significant color 
terms. Fig.~\ref{color} shows the de-extincted magnitude differences plotted 
with the $J - H$ colors from \citet{g03}. No significant color terms are apparent 
by eye and the color term slopes with errors are consistent with zero 
(Table~\ref{phot}). The reddest standards could not be observed in K after a 
computer disk unexpectedly reached capacity late in the night of 2006 May 12.
We do not feel that the slight positive correlation for K evident 
in Fig.~\ref{color} justifies the inclusion of a K color term.  
We expected a negligible color term as the filters were originally 
selected by considering their transmission and the quantum 
efficiency of the Astrocam detector to minimize any color term 
(A. Henden, private communication). Table~\ref{phot} summarizes the 
extinction and color term results, and the increase to the photometric errors 
applied in Fig.~\ref{ext}. Our extinction coefficients are in reasonable 
agreement to those determined by \citet{g03}. 

The J data in Fig.~\ref{ext} exhibit a larger scatter than the other filters. 
This could be due to a variable water vapor content. Although water 
is not the only source of opacity it is a significant effect and is more 
pronounced in the J filter compared to H and K \citep{mb79,vcm89}.
Fortunately a ground based Global Positioning System (GPS) meteorology 
station is located on the Camp Navajo Army Depot 8.3 km WNW from the USNO-Flagstaff
Station . This station (station code FST1) is part of the National Oceanic and 
Atmospheric Administration GPS Integrated Precipitable Water (IPW) Vapor (NOAA GPS$-$IPW) 
network. The system is described in detail in \citet{wg00}. The IPW content 
in centimeters over the site is recorded every 30 minutes, and the data 
are available via the world wide web\footnote{Observations courtesy 
of the NOAA Earth Systems Research Laboratory: http://gpsmet.noaa.gov}.

\begin{figure}
\epsscale{1.0}
\plotone{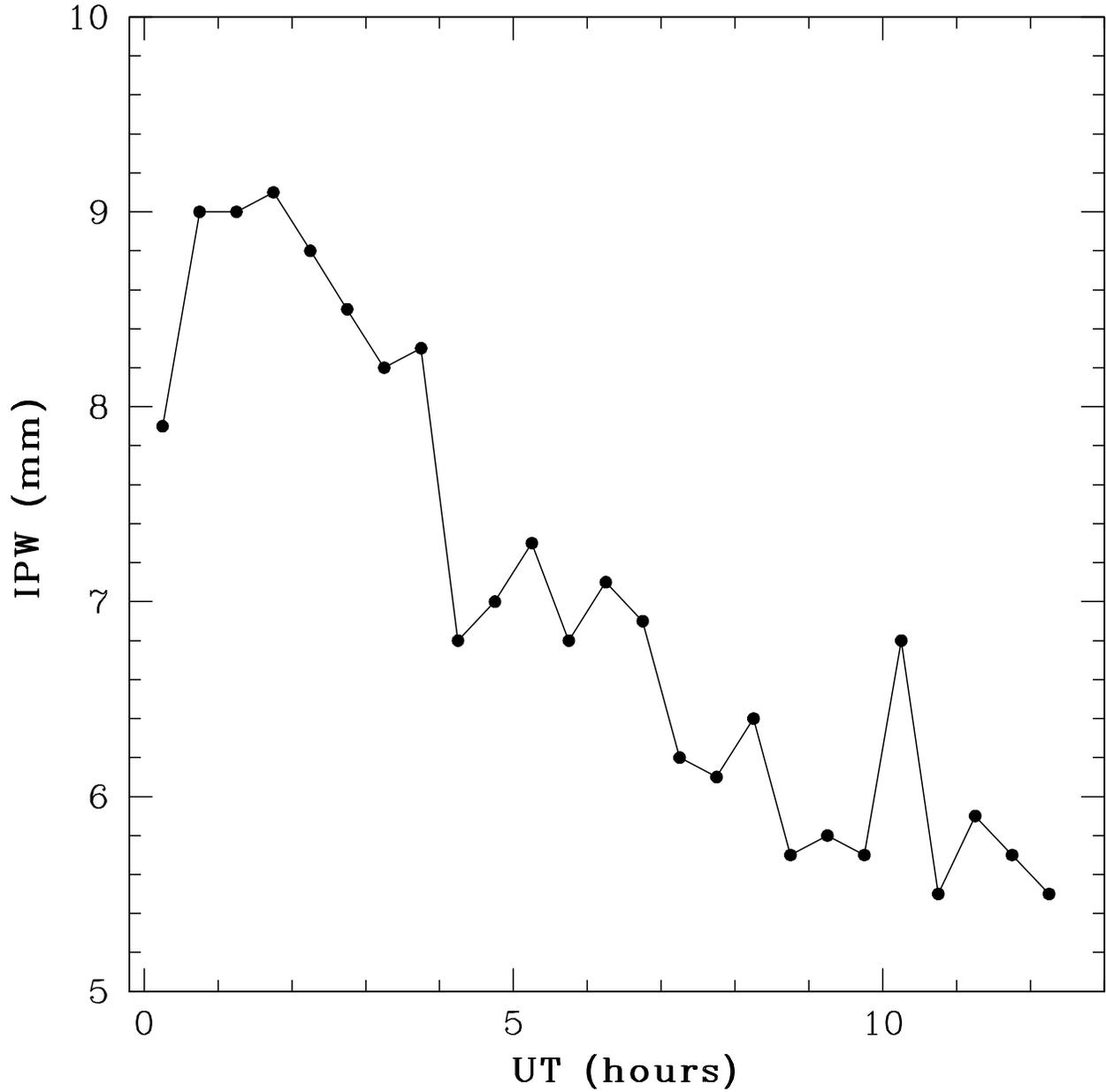}
\caption{The integrated precipitable water (IPW) content measured at the 
Camp Navajo Army Depot ground based GPS receiver on 2006 May 12.
Please note the disclaimer at the end of this paper. \label{ipw}}
\end{figure}

\begin{figure}
\epsscale{1.0}
\plotone{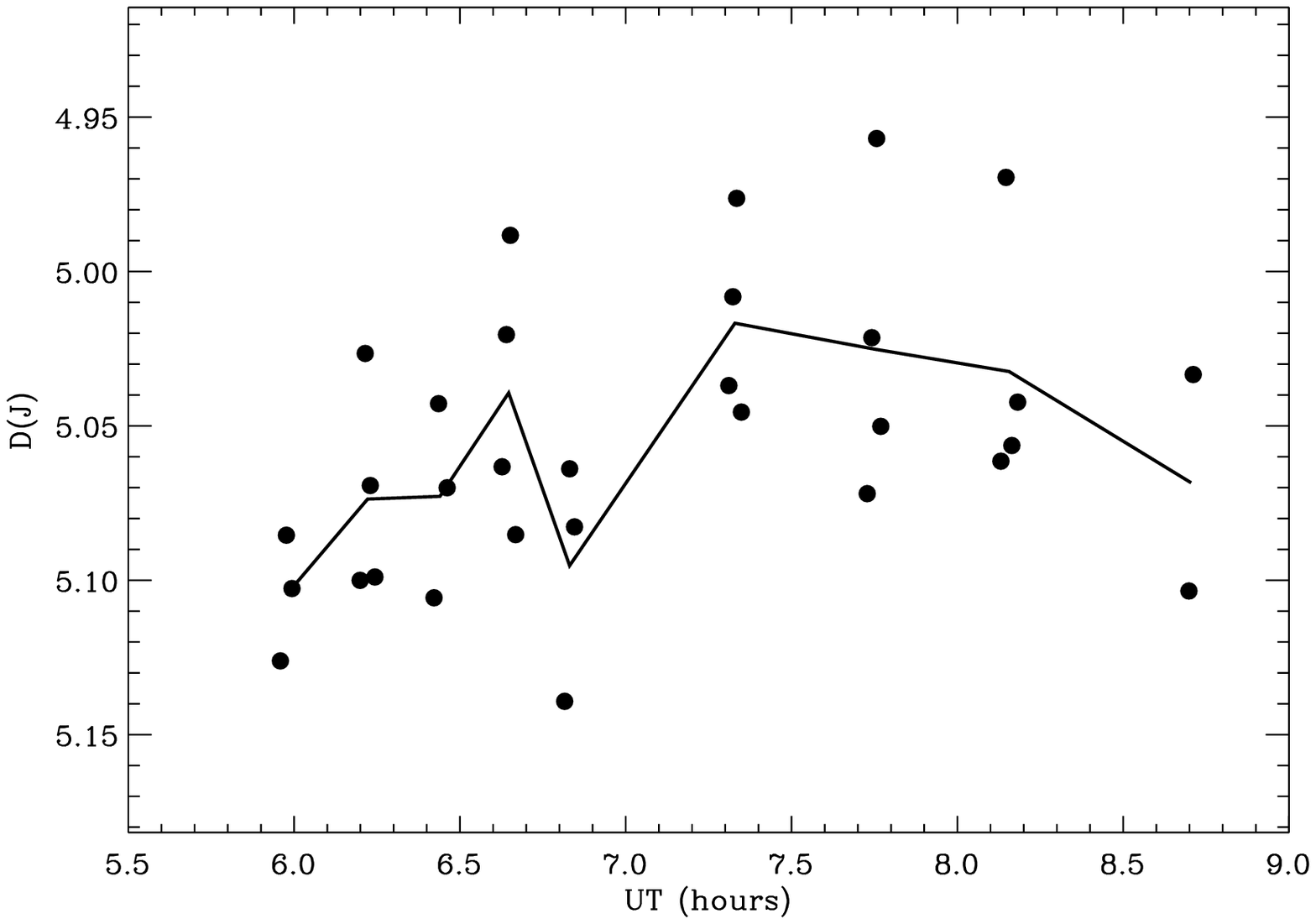}
\caption{The de-extincted J magnitudes of Fig.~\ref{color} as a 
function of UT time. Error bars are excluded for clarity. The solid 
line shows the average of the de-extincted magnitudes versus time to 
guide the eye for any trend correlated with Fig.~\ref{ipw}. \label{dj_time}}
\end{figure}

\begin{figure}
\epsscale{1.0}
\plotone{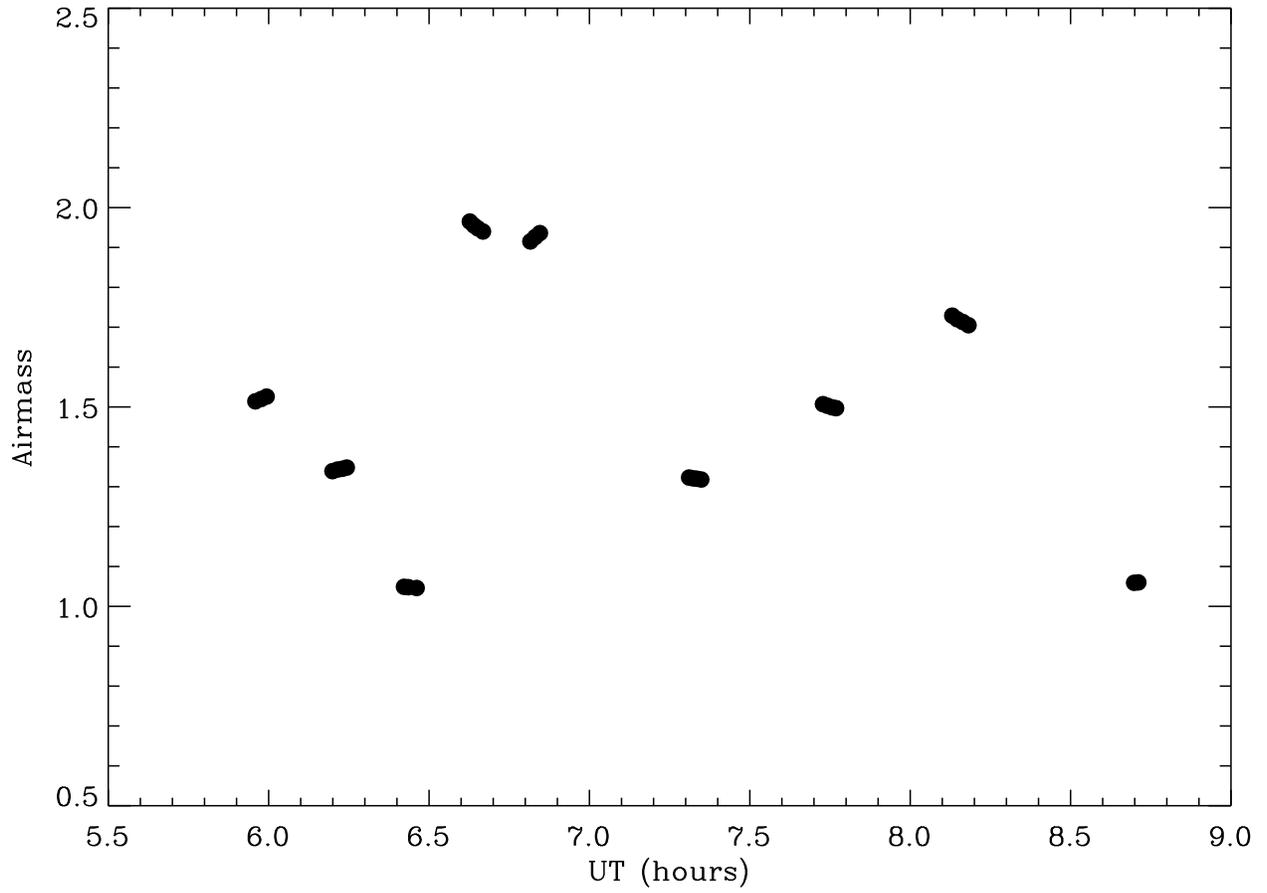}
\caption{The airmass of the observed de-extincted J magnitudes of Fig.~\ref{dj_time} 
as a function of UT time. \label{amass}}
\end{figure}

Our IR observations of $\delta$ Sco and the standards were undertaken 
between 0557 UT and 0853 UT. During this time the IPW content 
decreased from 7 to 6 mm (Fig.~\ref{ipw}) at the Camp Navajo station. 
While it is beyond the scope of this work to quantify the results of 
this decrease we can qualitatively expect a smooth decrease to 
the opacity over time based on the GPS IPW monitoring. 
To explore the effects of a variable water 
vapor content we show the de-extincted J magnitudes as a function of UT 
time in Fig.~\ref{dj_time}. We also display in this figure a solid line 
which shows the average of the de-extincted magnitudes as a guide to 
identify any correlation with the water vapor content. It is also useful to 
consider Fig.~\ref{dj_time} in conjunction with the airmass of the 
observations which we show in Fig.~\ref{amass}. For example if all 
the high airmass standard star observations occurred early or late in
the night they may be biased by a variable opacity and significantly 
effect the linear extinction fits. 

Considering Fig.~\ref{dj_time} a slight brightening may occur from 6 
to 8 hours UT, qualitatively consistent with a decreasing water vapor 
content. Furthermore, our 2 observations near an airmass of 2 are closely 
spaced in time but after correcting for extinction they differ in 
Fig.~\ref{dj_time} by approximately 0.05 magnitudes. The 
scatter in the standard star magnitudes in the individual sets of 
dithered images would, if due to a variable opacity, suggest a 
rather high frequency variable water vapor signal. Although the 
time sampling in Fig.~\ref{ipw} might preclude detection of such a signal 
some large 1 mm IPW swings should have been apparent even with the 
30 minute sampling interval. While the general trend in Fig.~\ref{dj_time} 
may be indicative of a changing water vapor content we cannot conclusively 
explain the scatter within the individual sets of dithered images except to 
invoke a problem in the sky subtraction. This we noted earlier when we 
saw the divergence in the curves of growth beyond a radius of 50 pixels 
(Fig.~\ref{cog}). Knowledge of the water vapor content 
can be used with a model of total atmospheric extinction to calibrate 
out extinction effects \citep{angione} with the aid of a radiometer. 
The proliferation of GPS-IPW monitoring sites and their provision near 
astronomical observatories could be used to calibrate out effects of 
atmospheric extinction \citep{gut} and it will be interesting to explore 
this possibility in the future. 
  
\begin{figure}
\epsscale{1.0}
\plotone{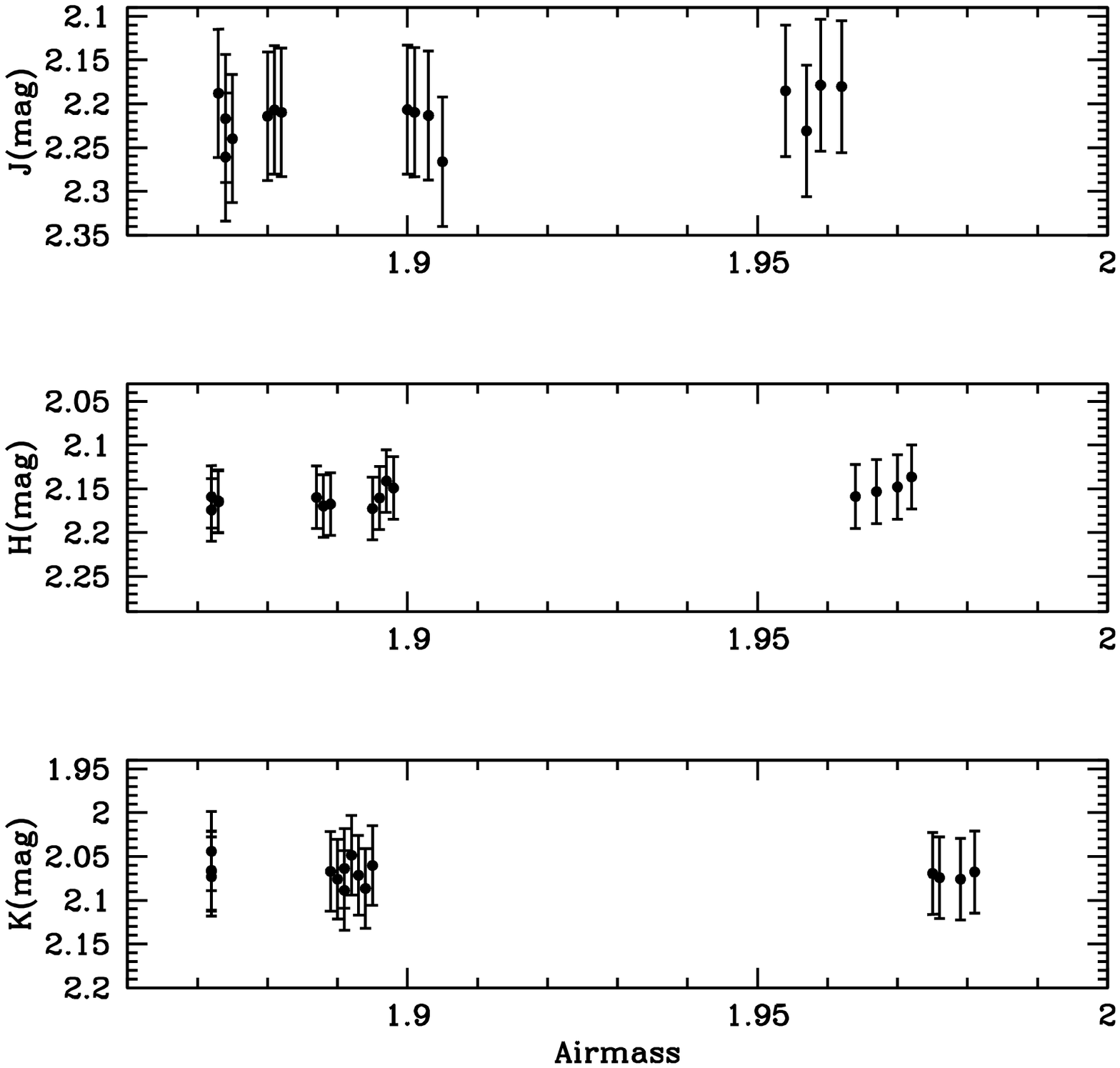}
\caption{Using the photometric solution for extinction in Table~\ref{phot} 
we determined the individual $\delta$ Sco J, H, and K magnitudes. These 
are shown above as a function of airmass. The averages of these points 
appear in our final solution for $\delta$ Sco in Table~\ref{dsco-tbl}. \label{dsco-fig}}
\end{figure}

Using only the extinction terms in Table~\ref{phot} we determined 
the magnitudes of $\delta$ Sco for 46 images. Fig.~\ref{dsco-fig} 
shows these individual measurements which within the errors appear flat 
as a function of airmass. The errors are dominated by the errors in the 
extinction terms due to the large counts in the $\delta$ Sco frames. 
We averaged the data and reduced the errors by $\sqrt{N}$ and found 
formal errors of 2\% or less in the individual magnitudes. Examination 
of Columns 9 and 10 of Table 1 from \citet{g03} allows us to 
establish that a lower level floor of 3\% is a more probable 
uncertainty for the individual magnitudes based on their many 
nights of observations at the Flagstaff Station. We therefore 
adopt this as our final uncertainty in the individual 
magnitudes, and propagate this in quadrature to produce 
the uncertainties in the colors of $\delta$ Sco.   
Table~\ref{dsco-tbl} lists the magnitudes and colors with 
this final uncertainty estimate. The magnitudes in Table 3 are identical within the errors to those of \citet{car06} observed 10 months earlier.

\section{Theoretical Be Star Models}

Several models of the thermal structure of Be star disks exist that can reproduce the spectral energy distribution of a star with its surrounding material. The theoretical models produced here were constructed with BEDISK, a non-LTE code that predicts the thermal structure of the disk using a realistic chemical composition. The principal features of BEDISK are discussed in detail in \citet{sig07}. We present here a brief overview of the code relevant to this investigation.

\subsection{Modeling a Be Star with BEDISK} 

The stellar radiation that reaches any particular point in the disk of the Be star depends on the emissivity and the opacity of the gas present throughout the entire disk. Thus, the photoionizing radiation field at a particular location in the disk is dependent on the density, temperature and atomic level populations at other locations in the disk. For a given photoionizing radiation field at each location in the circumstellar disk, the code solves the equations of statistical equilibrium (non local thermodynamic equilibrium) for the level populations for each atom and each ion included in the computation. Having produced the atomic level populations, the rates of heating and cooling due to the various atomic processes in the material can be computed at each disk location. Since the disk is assumed to be in radiative equilibrium, the temperature at each disk location can be iteratively obtained by balancing the rates of energy gain and energy loss. As the temperature profile is adjusted, the density profile is also adjusted in order to maintain vertical, hydrostatic equilibrium throughout the disk.

It is assumed that the radial density distribution of the disk falls off as a power law in the equatorial plane described as: 
\begin{equation}
\rho(R) = \rho_{o}(R/R_*)^{-n}
\end{equation}
where $\rho_o$ represents the density of the circumstellar material closest to the central star and $R \geq R_*$. This power law parameterization of the circumstellar disk follows the prescription of \citet{wat86}, \citet{cot87}, and \citet{wat87}, who constructed models to match the IR excess of Be stars. A second assumption is that the circumstellar gas is in vertical isothermal equilibrium perpendicular to the plane of the disk. The code accepts a user defined set of atomic models which contain a list of the energy levels and transitions for each atom and ion included in the disk model. The BEDISK code incorporates 425 levels and over 1000 radiative and collisional transitions for nine elements and thirty-one ions \citep[see Table 1 in][]{sig07}. 

\subsection{H$\alpha$ Emission Line}

The $JHK$ photometric observations of $\delta$~Sco were obtained on
2006 May 12. To measure the shape and overall strength of the
H$\alpha$ emission line close in time to the IR observations, we have
obtained spectra of $\delta$~Sco on 2006 April 19 and 2006~May~22.  The spectroscopic
observations were obtained using a fiber-fed echelle spectrograph
connected to the John S.~Hall 1.1~m telescope at the Lowell
Observatory.  The spectra in the H$\alpha$ line region were obtained
at a resolving power of $\sim$10,000 and with pixel spacing of 0.022 
nm (resulting in 3 pixels per resolution element). The signal-to-noise ratio (SNR) in the 
continuum level near H$\alpha$ line is approximately 100 for both 
spectra, and therefore the SNR of the emission line is around a few 
hundred. The spectroscopic observations were reduced using
standard reduction routines written in IDL\footnote{Interactive Data
Language of RSI, ITT Industries, Boulder, CO}~\citep{Hall94}, and the
resulting H$\alpha$ profiles of $\delta$~Sco obtained on two different nights in 2006 are shown in Figure~9. We have not corrected for the effects of telluric lines, which can be
seen in the Figure as shallow absorption lines, however, we have
reduced their effect on the equivalent width~(EW) measurement of the
H$\alpha$ emission line by excluding the contributions from regions
that go below the continuum level near the emission line. The
resulting EW measurements for $\delta$~Sco are -20.0 $\pm$ 0.06$\;$\AA\ on 2006~April~19 and -13.8 $\pm$ 0.04$\;$\AA\ on 2006~May~22. The uncertainties associated with our EW measures are estimated based on the limitation of the continuum level determination (which we estimate to be around 3\%) and are not associated with the SNR of the continuum. \citet{vol06} discuss the error analysis in equivalent widths in the case of photon noise limited case, and our uncertainties are equivalent to their analysis when the uncertainty associated with the continuum estimation dominates (i.e., when the second term in their eq. 5 is the main contributor to the uncertainty).

The model H$\alpha$ profiles were determined by formally solving the transfer equation along a set of rays through the system viewed at an  angle of $i$ relative to the line-of-sight ($i=0^o$ is pole-on,  
and $i=90^o$ is equator-on). We adopt a value of $i = 30^o$ to compute the profiles. This $i$ is in agreement with the values quoted by both \citet{car06} and \citet{mir01}. The disk is assumed to be in pure Keplerian rotation, and the equatorial rotational velocity of the star was chosen so that the measured $v\sin i$ of $~ 150$ km/s \citep{mir01} of the star was consistent with $i = 30^o$. For rays terminating on the stellar surface, a photospheric H$\alpha$ line profile is computed using the {\sc synthe} code and the Stark Broadening routines of \citet{bar03}; these profiles are based on the hydrogen populations computed for the appropriate LTE, line-blanketed model atmosphere adopted from \citet{kur93}. In the disk, the hydrogen emissivity and opacity followed directly from the hydrogen 
level populations determined using the self-consistent thermal structure. As a result, the full dependence of the H$\alpha$ line emissivity and opacity on the physical conditions throughout the disk were retained. The Stark broadening routines of \citet{bar03} were also used to compute the local H$\alpha$ line profile throughout the disk. This is important as these routines can handle high disk densities for which the assumption of a simple Gaussian or Voigt profile for the H$\alpha$ line would be invalid. The H$\alpha$ line profile was obtained by summing over the rays weighted by their projected area on the sky. Finally, the profile was convolved with a Gaussian having FWHM of 0.656 \AA ~so that the predicted profile matched the resolving power of $10^4$ of the spectroscopic observations.

\subsection{Modeling $\delta$ Scorpii}
We adopted the stellar parameters chosen by \citet{car06} in their study of the properties of $\delta$ Sco's circumstellar disk. The radius and mass of the star are 7.0 $R_\odot$ and 14.0 $M_\odot$, respectively, and the star's effective temperature is 27 000 K. We constructed models for seven values of $n$ ($n$ = 1.5, 2.0, 2.5, 3.0, 3.5, 4.0 and 4.5) with varying values of $\rho_o$ beginning with 1.0 $\times$ 10$^{-12}$ g cm$^{-3}$ and increasing the density up to 5.0 $\times$ 10$^{-10}$ g cm$^{-3}$. We then integrated the predicted spectral energy distribution of the stellar models over the J, H and K bandwidths in order to calculate the theoretical J-H and H-K colors. The results of our work are plotted in the color-color diagram illustrated in Fig.~\ref{f10}. 

\begin{figure}
\plotone{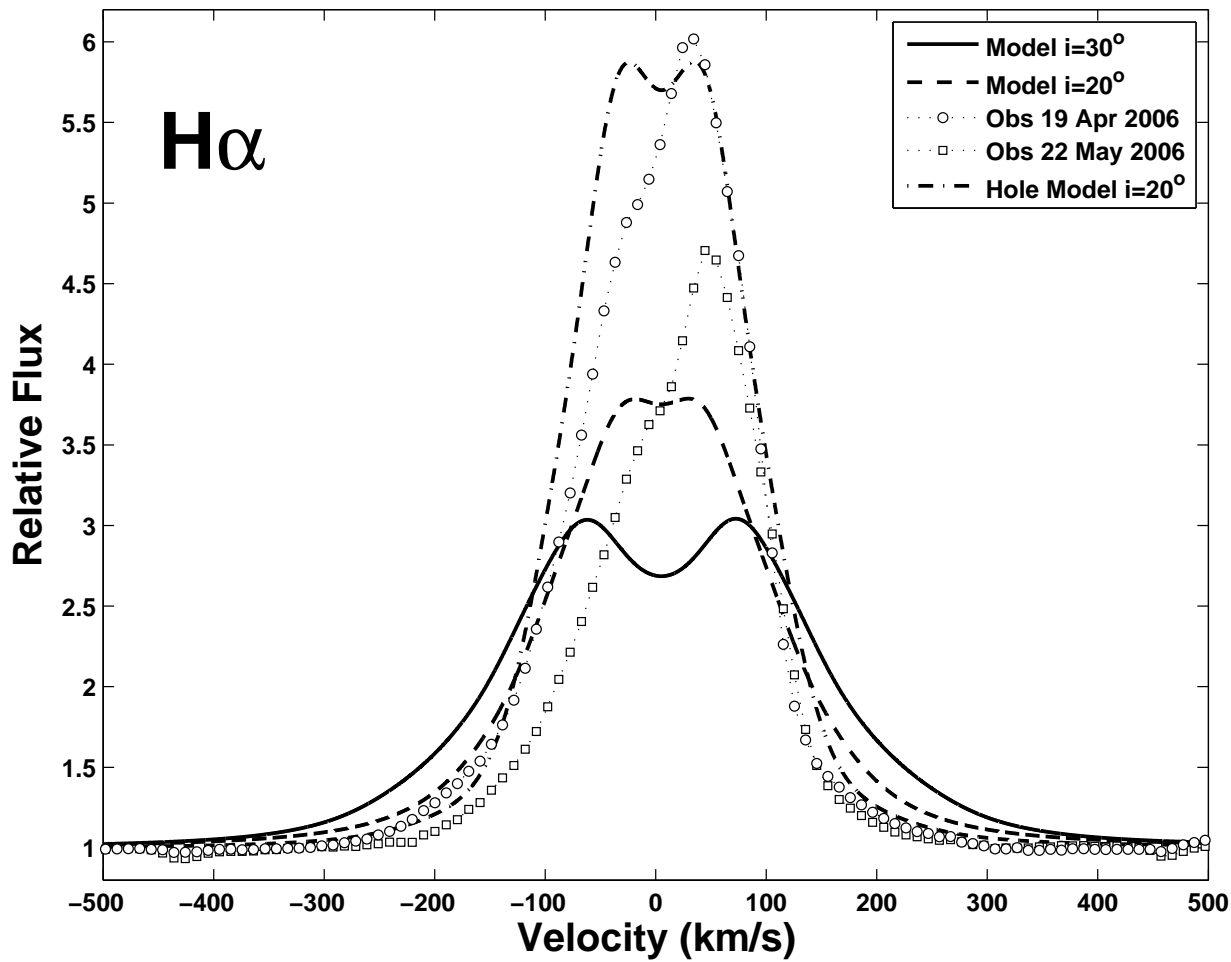}
\caption{Comparison of the observed H$\alpha$ lines taken on 19 April 2006 (dotted line with circles) and 22 May 2006 (dotted line with squares) and the theoretical lines produced by the model with disk parameters n = 4.0 and $\rho_0$ =5.0 $\times$ 10$^{-10}$ g cm$^{-3}$  and with a inclination of $i = 30^o$ (solid line) and $i = 20^o$ (dashed solid line).  We also show a model with an evacuated annulus between 1.5 and 4.0 stellar radii which had a density of 0.01 compared to our single power-law in this region (dot-dashed line).\label{f9}}
\end{figure}

\begin{figure}
\plotone{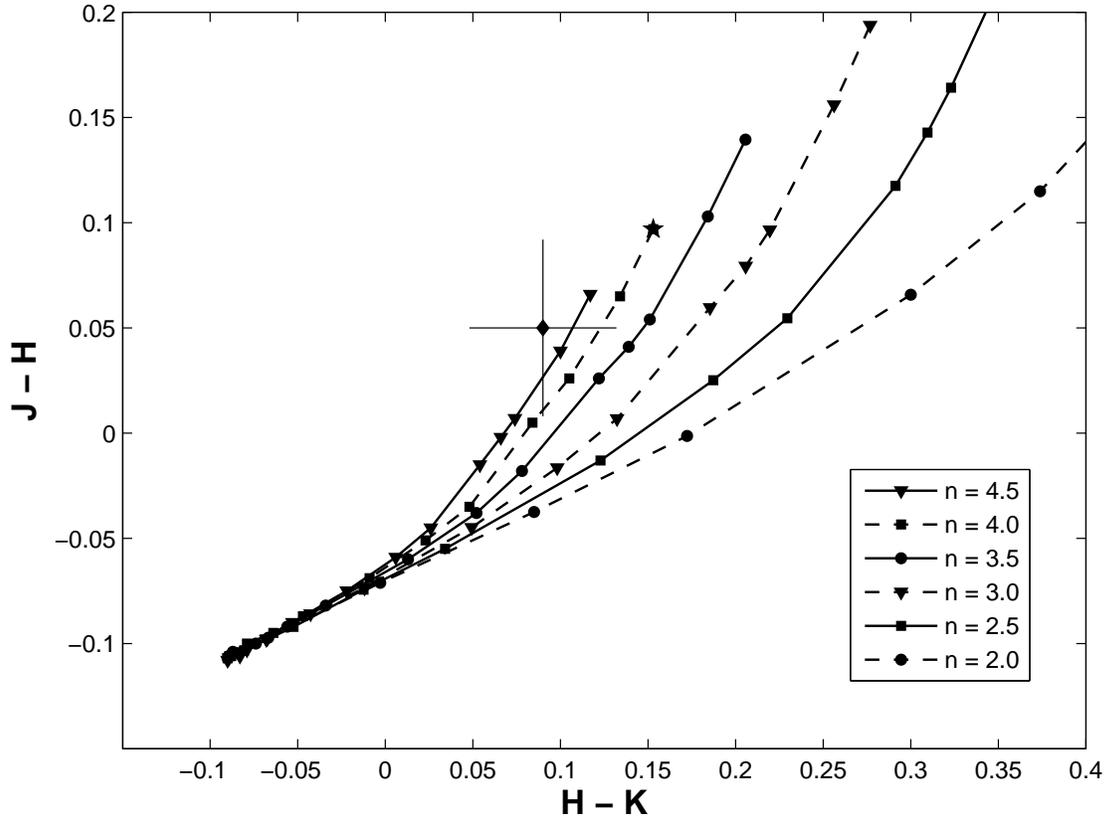}
\caption{Color-color diagram for the constructed models. The single point represents the combined photometric observations described in section 2. Each curve represents a different power-law density exponent with disk-base densities decreasing from left to right. The star shows the best-fit disk parameters obtained by constraining the solutions with spectroscopy. \label{f10}}
\end{figure}

In order to determine our best-fit model, we computed theoretical
H$\alpha$ profiles for a subset of models with synthetic colours to the observed value. We compared the equivalent widths and line
shapes of these predictions to H$\alpha$ spectra bracketing the
IR observations in time. These models corresponded to a range of $n$
from 3.5 to 4.5 and $\rho_o$ from 5.0 $\times$ 10$^{-11}$ to 5.0 $\times$
10$^{-10}$ g cm$^{-3}$. The models with the larger values of $n$
have a greater drop in density with radial distance from the star and
therefore correspond to the larger values of $\rho_o$ quoted. Within
this subset, the theoretical equivalent widths range from a minimum
of 7.1$\;$\AA\ to a maximum of 18.8$\;$\AA\ for $i=30^o$. Changing the
value of $i$ by $\pm\,\sim\,10^o$ for this subset of models changes the
equivalent width by less than $\sim\,1\;$\AA. However, changes in $i$
do significantly alter the shape of the profile. Furthermore, we produced our emission line profiles with the inclusion of a hydrostatic subroutine that enables our code to produce fully self-consistent disk densities and temperatures. Fig.~\ref{f9} compares the best-fit predicted H$\alpha$ line profile to the observed lines. Note that the observed line changes dramatically in strength and shape between the two bracketing observations taken about one month apart. This indicates that there are ongoing changes in the disk density structure during this short time period. 

As can be seen from the figure, the computed line profile for the $i=30^o$ model is too wide in the
wings as compared to the observations. To attempt to reduce the width of the line, we have computed the H$\alpha$ line profile for a slightly smaller inclination of the disk relative to the line of sight ($i=20^o$) and also added an evacuated annulus between 1.5 and 4.0 stellar radii which had a density of 0.01 compared to our single power-law in this region. We acknowledge that the smaller inclination of $i=20^o$ seems inconsistent with the orbital solution for $\delta$ Sco \citep{mir01} which gives the orientation of the orbital plane relative to the line of sight as $38^o\pm5^o$, assuming that the disk lies in the orbital plane of the system. The smaller inclination angle and the addition of the evacuated annulus do reduce the width of the line with the $i=20^o$ hole model coming closest to the observations. Nevertheless, we are not proposing the more complex hole model as the best fit as there are a number of parameters in this model (location of annulus and density contrast in the annulus)
which would need additional observational constraints. In addition, we note that the observed lines show a definite asymmetry which cannot be modeled by our axisymmetric calculations. Further study of the line asymmetry could reveal a dynamic envelope density structure such as a one-armed oscillation in the equatorial disk \citep{vak98}. In the future, we plan on incorporating density perturbation into our disk models so that we can replicate the asymmetric line profiles. 

\begin{figure}
\plotone{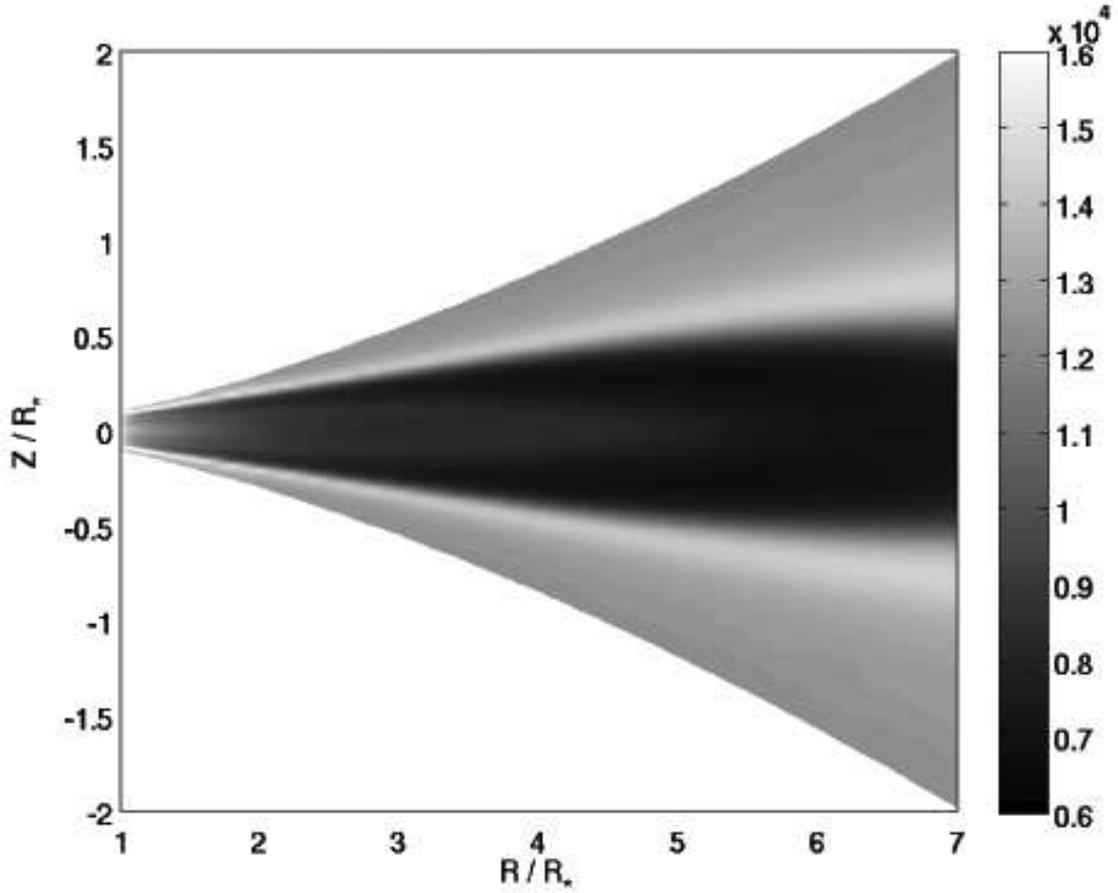}
\caption{Temperature distribution of the disk with model parameters n = 4.0 and $\rho_0$ =5.0 $\times$ 10$^{-10}$ g cm$^{-3}$. We obtain a minumum temperature of 6700 K in the cool, dense region along the equatorial plane of the star and an average overall disk temperature of 12000 K.\label{f11}}
\end{figure}

\begin{figure}
\plotone{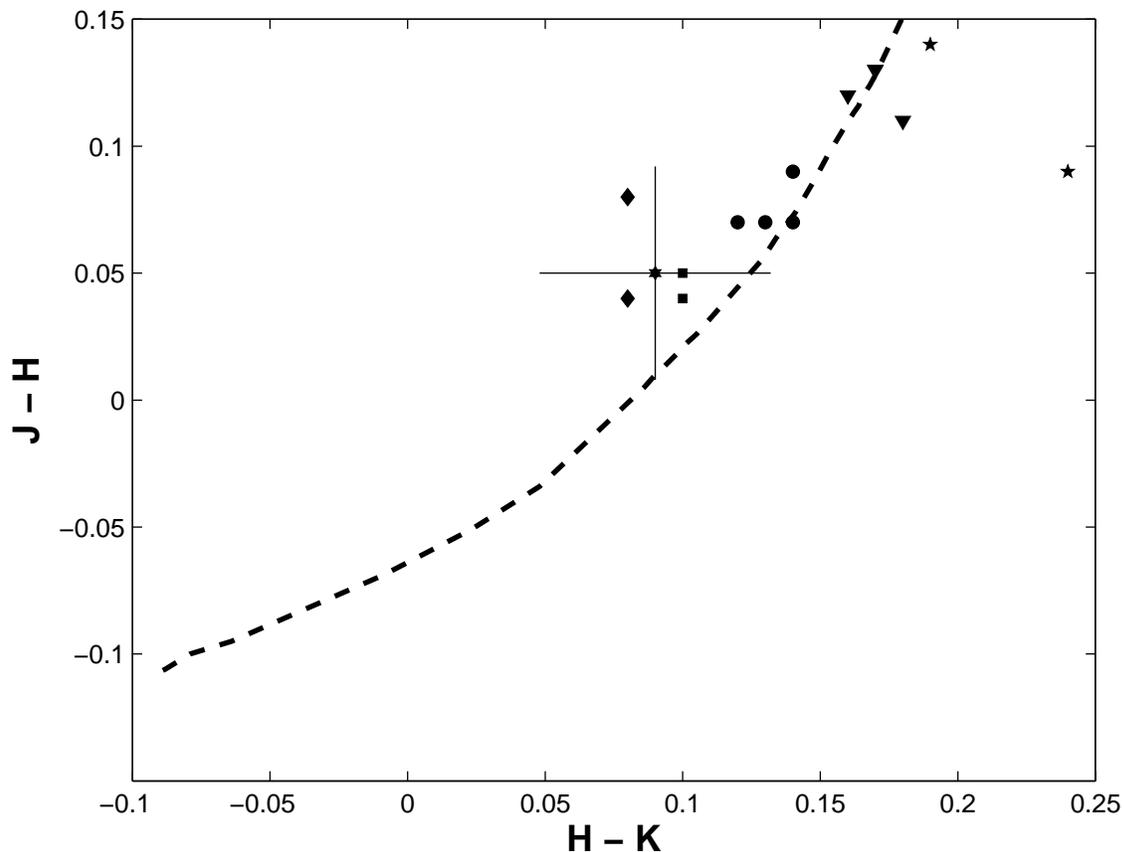}
\caption{JHK colours from \citet{car06} plotted against the observation described in section 2 (point with error bars) and one of the model curves from Figure 9 for a fixed n = 4.0 and a disk-base density varying from $\rho_o$ = 1.0 $\times$ 10$^{-12}$ g cm$^{-3}$ to $\rho_o$ = 1.0 $\times$ 10$^{-9}$ g cm$^{-3}$ (dashed line). The JHK colours are grouped together based on time of observation (ranging from 2000-2005) and location. The earliest observations are the diamonds, followed by the circles, the stars, the triangles and the squares.\label{f12}}
\end{figure}

In comparing our work with the study on $\delta$ Sco presented by \citet{car06}, we emphasize our agreement with some of their results. First, we note that the disk temperature profile for our best-fit parameters (Fig.~\ref{f11}) shows reasonable agreement with the average temperature distribution of their best-fitting model. Both theoretical models predict a cool dense region along the equatorial plane near the star.  We obtain a minimum temperature of 6700 K in agreement with the minimum temperature quoted by \citet{car06} of 7000 K. The average overall temperature we obtain is 12000 K, slightly less than the value of 16000K of \citet{car06}. We plan on performing a detailed comparison of our computational code with Carciofi et al.'s non-LTE Monte Carlo radiation transfer code in a future work. Furthermore, our predicted colours of the star from our set of models cover the range of values in which we find the photometric observations presented in \citet{car06} (Fig.~\ref{f12}). The observations, taken over a five year period during which the disk surrounding $\delta$ Sco was developing, suggest that our radial drop-off model for the structure of the disk is a reasonable approximation for the inner portion of the disk where H$\alpha$ and the near IR continuum form. Notice that our theoretical models constructed with $n = 4.0$ for a range of $\rho_o$ fall within the locus of the Carciofi et al. (2006) photometric observations.

\subsection{Binary Component}

\begin{figure}
\plotone{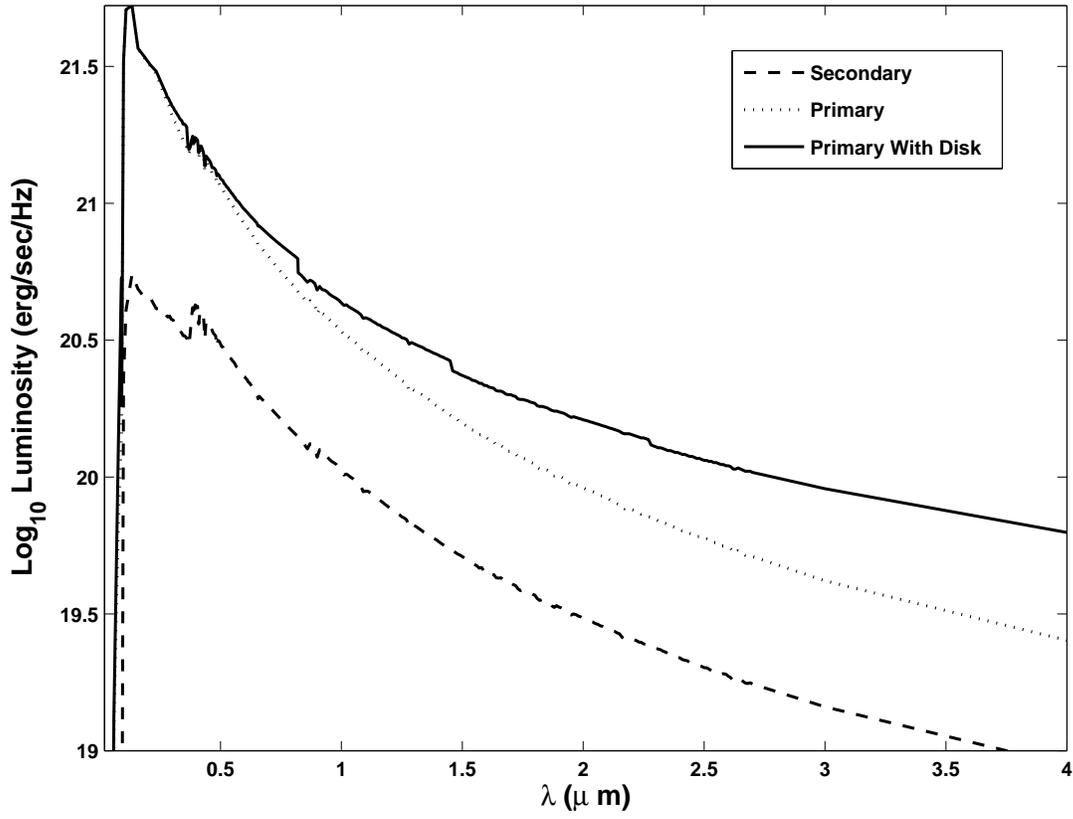}
\caption{Spectral energy distribution for the secondary star, the isolated primary star and the primary star surrounded by a circumstellar disk with parameters n = 4.0 and $\rho_0$ =5.0 $\times$ 10$^{-10}$ g cm$^{-3}$.\label{f13}}
\end{figure}

In order to ensure that our predicted IR excess for the disk of $\delta$ Sco is an accurate measure,  we took into consideration the binarity of the system by assessing the impact of the secondary on the observed infrared continuum. We set the spectral type of the secondary star through its optical magnitude difference from the primary. \citet{bed93} found that the secondary was around 1.5 mag fainter than the diskless B0.2 IV primary. Our estimate of a B3 spectral type for the secondary is supported by comparing the estimated mass of the object \citep[the mass ratio of the system is 1.5 - 2 according to][]{bed93} to the values listed in \citet{all99}. Assuming this spectral type for the secondary, we re-calculated our predicted colours by including the SED from the secondary star. Fig.~\ref{f13} shows the spectral energy distribution of the secondary, the primary and the primary with its circumstellar disk.

The inclusion of the secondary star did not change the general range of the power-density exponents. For each value of $n$, we found that this approximation did not shift the value of our best-fit disk-base density by more than 0.03 mag in H-K and 0.02 mag in J-H. While the binarity of the system does affect the observed infrared continuum, our investigation shows that the influence of the secondary is not considerable enough to significantly alter the range of parameters that match the infrared excess observation.

\section{Conclusion}

In summary, we found that $\delta$ Sco's J, H and K magnitudes are identical within the errors to those observed 10 months earlier by \citet{car06}. We modeled $\delta$ Sco's circumstellar disk and compared the theoretical infrared excess to photometric observations of the excess. These observations, combined with contemporaenous spectroscopic observations, are particularly important for studying Be stars in their active phases \citep{car06}. We used a non-LTE computational code to model the thin equatorial disk of gas surrounding $\delta$ Sco, employing a simple power-law for the density distribution of disk material. Although the stellar activity suggests a more complicated density distribution in the circumstellar disk, monitoring the infrared excess and determining the average value of the power-law index remains an important part of studying $\delta$ Sco \citep{oka07}. With this in mind, we calculated the spectral energy distribution of the star and its disk and we derived the theoretical infrared color excesses. We compared these predictions to our infrared photometric observations of $\delta$ Sco and further constrained the disk parameters by modeling the H$\alpha$ emission line and comparing it to spectra taken before and after the time of the photometric observation. Our best fit parameters for the circumstellar disk were found to be $\rho_o$ = 5.0 $\times$ 10$^{-10}$ g cm$^{-3}$ with a power-law density exponent of $n$ = 4.0. Overall, we found acceptable parameters within the range for $n$ from 3.5 to 4.5 and $\rho_o$ from 5.0 $\times$ 10$^{-11}$ to 5.0 $\times$
10$^{-10}$ g cm$^{-3}$. Since $\delta$ Sco's disk changes dynamically, our analysis places constraints on the disk at the time of the observations. It will be interesting to see if this value for the power-law index changes with time which will provide evidence to assess the applicability of the viscous disk model. Recently several investigations have suggested specific detailed features in density distributions in Be star disks. For example, \citet{zor07} find that the power-law index is less than one within the first 3 stellar radii. This implies that the density distribution near the star is constant or increasing. \citet{riv01} proposed that Be star disks are composed of a series of rings.
This concept has been studied recently by by \citet{mei06} and \citet{car07}. Other proposed features in the density distribution include a rotating one-armed spiral density wave based on the variation of
the ratio of violet to red intensities in asymmetric double-peaked line profiles \citep[see, for example,][]{wis07}. For $\delta$ Sco specifically, \citet{car06} suggest that the disk geometry must change over time or that material from the stellar photosphere must be ejected at higher stellar latitudes to explain variations in brightness of the system. Models which include these types of density enhancements or depletions will necessarily include additional free parameters which will require further monitoring of $\delta$ Sco with contemporaneous observations. In the future, we plan to obtain other observables of $\delta$ Sco in order to place further constraints on the radial density distribution in the circumstellar disk of the star and to trace the evolution of the disk.

\acknowledgments

We thank the Lowell Observatory for the telescope time used to obtain the H$\alpha$ line spectrum. This work is supported by the Canadian Natural Sciences and Engineering Research Council through a graduate scholarship for RJH and research grants for CEJ and TAAS. NV's participation was funded by 
the Office of Naval Research through a Science and Engineering Apprenticeship Program internship 
administered by the American Society for Engineering Education. RTZ thanks Mike DiVittorio 
for helpful discussions about the optics of the Strand reflector. 

Disclaimer Notice for use of NOAA Data:

Data, display products, and documentation available on this computer are 
meant for research.  All data and documentation are considered part of the
public domain and are furnished "as is."  These data were collected with
the intent of using them for meteorological research.  The United States
government, its instrumentalities, officers, employees, and agents make no 
warranties, express or implied, as to the accuracy, quality, usefulness of
the data, products, and/or documentation for any purpose.  They cannot be held
responsible for any circumstances resulting from their use, unavailability,
or possible inaccuracy.  Earth System Research Laboratory (ESRL) will make
data and/or display products available in World Meteorological Organization
(WMO), ESRL, or other standard formats which may change at times.  ESRL
reserves the right to suspend or discontinue this service, or portions
thereof, at any time.

\clearpage

\begin{deluxetable}{lccccc}
\tablecaption{S{\sc tandard} S{\sc tars}\label{stndrd}}
\tablewidth{0pt}
\tablehead{ \colhead{Star} & \colhead{K} & \colhead{J$-$H} & \colhead{H$-$K} & \colhead{Exp.(sec)} 
& \colhead{Co-adds}}
\startdata
Gl 406 & 6.071 & 0.619 & 0.364 & 0.20 & 100 \\
HD 106965 & 7.312 & 0.048 & 0.015 & 0.50 & 50 \\
HD 136754 & 7.135 & 0.016 & 0.005 & 0.30 & 60 \\
SA 108$-$475 & 7.986 & 0.682 & 0.136 & 1.00 & 20 \\
HD 161961 & 7.304 & 0.063 & $-$0.011 & 1.00 & 20 \\
Gl 748 & 6.302 & 0.524 & 0.243 & 0.20 & 100 \\
\enddata
\end{deluxetable}
\clearpage

\begin{deluxetable}{cccccccc}
\tablecolumns{8}
\tablewidth{0pc}
\tablecaption{2006 May 12 P{\sc hotometric} S{\sc olution} \label{phot}}
\tablehead{
\colhead{} & \colhead{} & \multicolumn{3}{c}{Extinction} & \colhead{} & \multicolumn{2}{c}{Color Term} \\
\cline{3-5} \cline{7-8} \\
\colhead{Filter} & \colhead{\# points} & \colhead{Ext.}       & \colhead{Ext. 0} &
\colhead{Factor} & \colhead{} & \colhead{Color term} & \colhead{Color 0} \\
\colhead{(1)} &\colhead{(2)} &\colhead{(3)} &\colhead{(4)} &\colhead{(5)} &\colhead{} &\colhead{(6)} &\colhead{(7)}}   
\startdata
J & 31 & -0.12 $\pm$ 0.03 & -5.06 $\pm$ 0.05 & 5.26 & & -0.003 $\pm$ 0.023 &  5.057 $\pm$ 0.010 \\
H & 25 & -0.16 $\pm$ 0.02 & -4.31 $\pm$ 0.02 & 2.76 & &  0.004 $\pm$ 0.014 &  4.307 $\pm$ 0.006 \\
K & 26 & -0.11 $\pm$ 0.02 & -5.30 $\pm$ 0.03 & 3.11 & & -0.018 $\pm$ 0.017 &  5.310 $\pm$ 0.007 \\
\enddata
\tablecomments{Col. 1: Filter. Col. 2: Number of data points used. Col. 3: Extinction coefficient (mag airmass$^{-1}$). 
Col. 4: Extinction zeropoint and error (mag). Col. 5: Factor used to increase errors in Fig.~\ref{ext}. 
Col. 6: Color term and error (mag color$^{-1}$). 
Col. 7: Color term zeropoint and error (mag).}
\end{deluxetable} 
\clearpage

\begin{deluxetable}{cc}
\tablecaption{D{\sc elta} S{\sc co} P{\sc hotometric} S{\sc olution} \label{dsco-tbl}}
\tablewidth{0pt}
\tablehead{\colhead{Filter or Color} & \colhead{Result and Uncertainty}}
\startdata
J & 2.21 $\pm$ 0.03 \\
H & 2.16 $\pm$ 0.03 \\
K & 2.07 $\pm$ 0.03 \\
J$-$H & 0.05 $\pm$ 0.042 \\
H$-$K & 0.09 $\pm$ 0.042 \\
\enddata
\end{deluxetable}
\clearpage

\clearpage

\clearpage

\clearpage

\clearpage 

\clearpage 

\clearpage

\clearpage

\clearpage

\clearpage

\clearpage

\clearpage

\end{document}